\documentclass{iau} 
\usepackage{graphicx}
\usepackage{hyperref}
\usepackage{amsmath}

\newcommand{\sii}{\relax \ifmmode {\mbox S\,{\scshape ii}}\else S\,{\scshape ii}\fi}
\newcommand{\siii}{\relax \ifmmode {\mbox S\,{\scshape iii}}\else S\,{\scshape iii}\fi}
\newcommand{\nii}{\relax \ifmmode {\mbox N\,{\scshape ii}}\else N\,{\scshape ii}\fi}
\newcommand{\oi}{\relax \ifmmode {\mbox O\,{\scshape i}}\else O\,{\scshape i}\fi}
\newcommand{\oii}{\relax \ifmmode {\mbox O\,{\scshape ii}}\else O\,{\scshape ii}\fi}
\newcommand{\oiii}{\relax \ifmmode {\mbox O\,{\scshape iii}}\else O\,{\scshape iii}\fi}
\newcommand{\neiii}{\relax \ifmmode {\mbox Ne\,{\scshape iii}}\else Ne\,{\scshape iii}\fi}
\newcommand{\te}{T$_{\mathrm{e}}$}
\newcommand{\oiiin}{[\oiii]$_\mathrm{n}$}

\title[A new tool to derive chemical abundances in Type-2 AGN] 
{A new tool to derive chemical abundances in Type-2 Active Galactic Nuclei}

\author[Garc\'ia-Benito, P\'erez-Montero, Dors et al.]   
{Rub\'en Garc\'ia-Benito$^1$,
Enrique P\'erez-Montero$^1$,
O.\ L.\ Dors Jr.$^2$,\\
J.\ M.\ V\'\i lchez$^1$,
M.\ V.\ Cardaci$^{3,4}$,
\and G.\ F.\ H\"agele$^{3,4}$}

\affiliation{$^1$Instituto de Astrof\'isica de Andaluc\'ia, \\ Apartado de correos 3004,
E-18080 Granada, Spain \\ email: {\tt rgb@iaa.es \tt epm@iaa.es} \\[\affilskip]
$^2$ Universidade do Vale do Para\'iba, Av. Shishima Hifumi,\\
2911, Cep 12244-000, S\~ao Jos\'e dos Campos, SP, Brazil \\[\affilskip]
$^3$ Instituto de Astrof\'isica de La Plata (CONICET-UNLP), Argentina \\[\affilskip]
$^4$ Facultad de Ciencias Astron\'omicas y Geof\'{\i}sicas, \\Universidad Nacional de La Plata, 
Paseo del Bosque s/n, 1900 La Plata, Argentina}

\pubyear{2019}
\volume{356}  
\jname{Nuclear Activity in Galaxies Across Cosmic Time}
\editors{M. Povi\'c, P. Marziani, J. Masegosa, H. Netzer,\\ S. H. Negu, \& S. B. Tessema, eds.}
\begin{document}

\maketitle

\begin{abstract}
We present a new tool for the analysis of the optical emission lines of the gas in the Narrow Line 
Region (NLR) around Active Galactic Nuclei (AGNs). This new tool can be used in large samples 
of objects in a consistent way using different sets of optical emission-lines taking into the 
account possible variations from the O/H - N/O relation. The code compares certain observed 
emission-line ratios  
 with the predictions from a large grid of photoionization models calculated under the 
most usual conditions in the NLR of AGNs
to calculate the   total oxygen abundance, nitrogen-to-oxygen ratio and ionization parameter. 
We applied our method to a sample of Seyfert 2 galaxies with 
optical emission-line fluxes from the literature. Our results confirm the high metallicity of 
the objects of the sample and provide consistent values with the direct method. The usage of models 
to calculate precise ICFs is mandatory when only optical emission lines are available to derive 
chemical abundances using the direct method in NLRs of AGN.
\end{abstract}

\firstsection 
\section{Introduction}

The energetic radiation coming from the central black holes in galaxies is partially re-emitted by the surrounding gas as very 
bright emission lines which in turn can be used to derived the physical conditions  in these 
extreme regions. Since they can be observed up to very high redshifts, Active galactic Nuclei (AGNs) are 
thus a powerful source for the study of cosmic evolution of galaxies.

It is widely accepted (\cite[Ferland \& Netzer 1983]{ferland83}) that the main mechanism of the narrow-line 
region (NLR) in AGNs is photoionization. However, it is also known that the total metallicity derived 
using the direct method (i.e. the \te\ method) gives sub-solar metallicities in AGNs, 
as compared to the predictions from photoionization models. Using a sample of NLRs of AGNs, 
\cite{dors15} found that the \te-method using the optical lines underestimated the oxygen 
abundances by an averaged value of $\sim$0.8 dex as compared to calibrations based on 
photoionization models.

Models are, therefore, a powerful tool to interpret the observed lines and provide valuable information 
to study chemical abundances. In this work, we describe a new code based on photoionization 
models to derive chemical abundances in the NLR in AGNs.

\section{The code}
In \cite{epm19} we present a full description of a new code to derive the total oxygen abundance, 
nitrogen-to-oxygen ratio (N/O), and the ionization parameter (U) from the analysis of optical emission 
lines in the NLR of type-2 AGNs. The code is based on the well proven 
HII-CHI-MISTRY\footnote{Publicly available in the webpage \url{https://www.iaa.csic.es/~epm/HII-CHI-mistry.html}.} 
code (hereafter HCM, \cite[P\'erez-Montero 2014]{hcm14}) originally developed for the analysis of 
star-forming regions. The advantages of the code are: 
a) it can be applied to a large number of objects in an automatic way; 
b) all objects are analyzed in a consistent way regardless of the set of input emission lines; c) it provides uncertainties for all the estimated 
quantities; d) it provides an independent estimation of the N/O ratio; and e) it is consistent with 
the direct method.

The code uses a grid of 5\,865 photoionization models run with the code \cite{cloudy} v.17.01. The models 
cover a wide range of the parameters space with typical NLRs conditions 
(see \cite[P\'erez-Montero et al. 2019]{epm19} for further details). The spectral energy distribution 
(SED) is composed by two components: the Big Blue Bump at 1 Ryd and a power law with spectral index 
$\alpha_\mathrm{X} = -1$. The continuum between 2KeV and 2500 \AA\ is modeled by a power law with spectral index 
$\alpha_\mathrm{OX} = -0.8$. All models were calculated using a spherical geometry with a filling factor of 0.1, a standard dust-to-gas ratio and a
constant density of 500 particles per cm$^{-3}$. In addition we checked the effect of changing in the models the $\alpha$(ox) down to -1.2 and enhancing the
electron density up to 2\,000 cm$^{-3}$ but no noticeable changes were found in the calculation of the chemical abundances using the method described here. 
For more details on the results of these
comparison see  \cite{epm19}. 
The models cover the range of 12 + log(O/H) from 6.9 to 9.1 in bins of 0.1 dex. 
The N/O range goes from -2.0 to 0.0 in bins of 0.125 dex and log \textit{U} from -4.0 to -0.5 in bins 
of 0.25 dex. 

The code uses as input the reddening-corrected relative-to-H$\beta$ emission line intensities from 
[\oii] $\lambda$3727 \AA, [\neiii] $\lambda$3868 \AA, [\oiii] $\lambda$4363 \AA, 
[\oiii] $\lambda$5007 \AA, [\nii] $\lambda$6583 \AA, and [\sii] $\lambda$$\lambda$6717+6731 \AA\ 
with their corresponding errors. However, the code is adapted to provide also a solution in case one 
or several of these lines are not given. 

In short, the work-flow of the code is as follows. First, the code constrain the parameter space searching 
for N/O as a weighted mean over all models, using optical emission lines for similar excitation, such as 
the ratio [\nii]$\lambda$6583 / [\oii]$\lambda$3727 or [\nii]$\lambda$6583 /  [\sii]$\lambda\lambda$6717+6731.
These ratios do not show almost any dependence on excitation and therefore N/O 
can be calculated without any assumption about the ionization parameter. Using the uncertainties of 
all the input observed lines, the code calculates the error using a Monte Carlo simulation. 
A set of line ratios such as [\oiii]$\lambda$5007/[\oiii]$\lambda$4363,
[\nii]$\lambda$6583/H$\alpha$, 

\begin{align*}
{\rm R23} = \frac{[{\rm O\textsc{ii}}] \lambda 3727 + [{\rm O\textsc{iii}}] \lambda\lambda 4959 + 5007}{\rm H\beta},
\end{align*}

\begin{align*}
{\rm O3N2} = \log \left(\frac{[{\rm O\textsc{iii}}] \lambda5007}{\rm H\beta} \cdot \frac{\rm H\alpha}{[{\rm N\textsc{ii}}] \lambda6583} \right),
\end{align*}

or

\begin{align*}
{\rm O2Ne3} = \frac{[{\rm O\textsc{ii}}] \lambda 3727 + [{\rm Ne\textsc{iii}}] \lambda3868}{\rm H\beta}
\end{align*}

\noindent (depending on the availability of the observed lines) are used in a second 
iteration to sample a subset of models constrained to the N/O values previously 
calculated to obtain the oxygen abundance and the ionization parameter. 

\section{Results}

\subsection{The control sample}
No empirical derivation of chemical abundances (i.e. no abundances using the direct method) in the 
NLR of AGNs using optical emission lines are available in the literature.
Therefore, we use as a control sample the abundance estimations by \cite{dors17} obtained from detailed 
tailored photoionization models using the \textsc{cloudy} code. They compiled a sample of 47 Seyfert 
1.9 and 2 galaxies at a redshift $z$ $\leq$ 0.1 providing the most prominent optical emission lines, 
including the auroral line [OIII] 4363 \AA. They do not provide an error estimation of the oxygen 
abundances obtained from their models.

\subsection{Comparisons}

\begin{figure}[h]
\begin{center}
\includegraphics[width=0.49\textwidth]{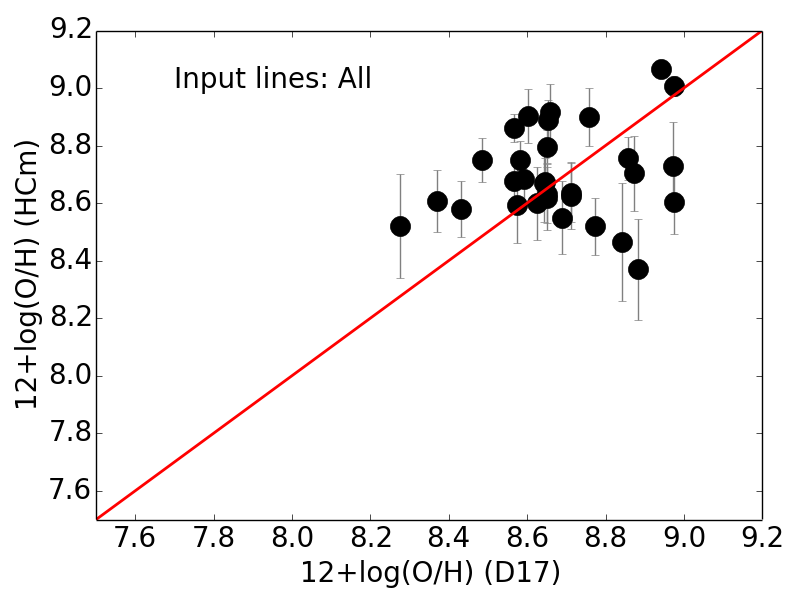} 
\includegraphics[width=0.49\textwidth]{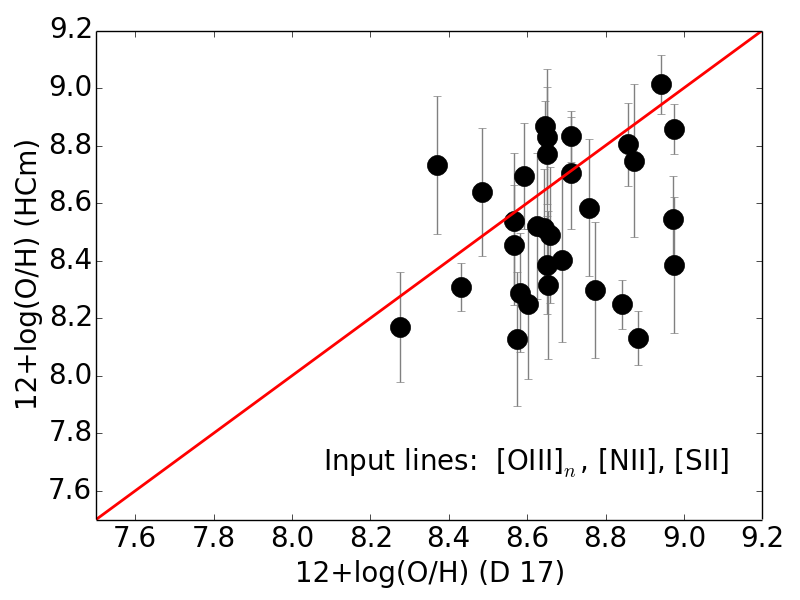}\\
\includegraphics[width=0.49\textwidth]{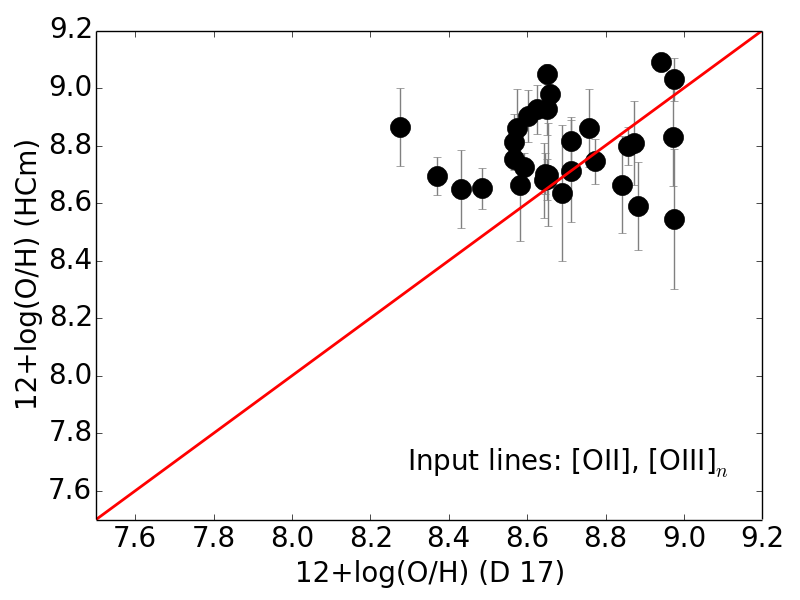} 
\includegraphics[width=0.49\textwidth]{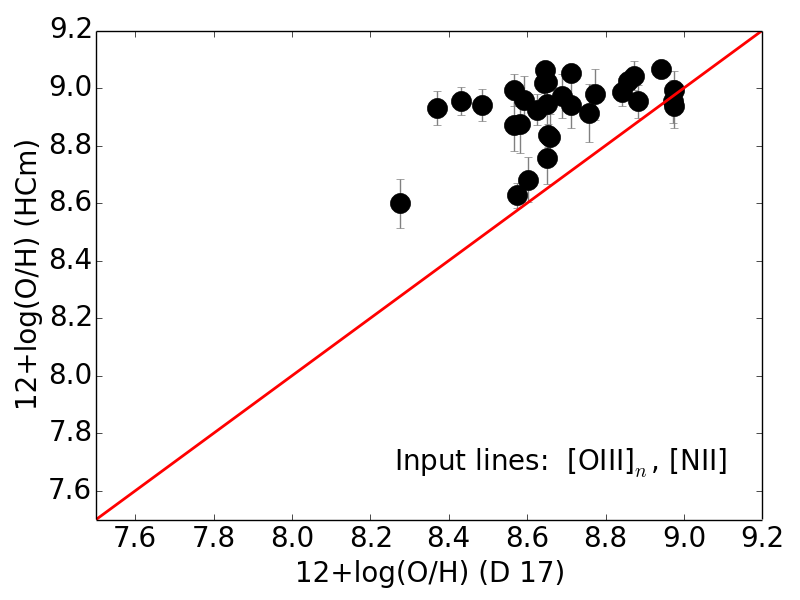}\\
\includegraphics[width=0.49\textwidth]{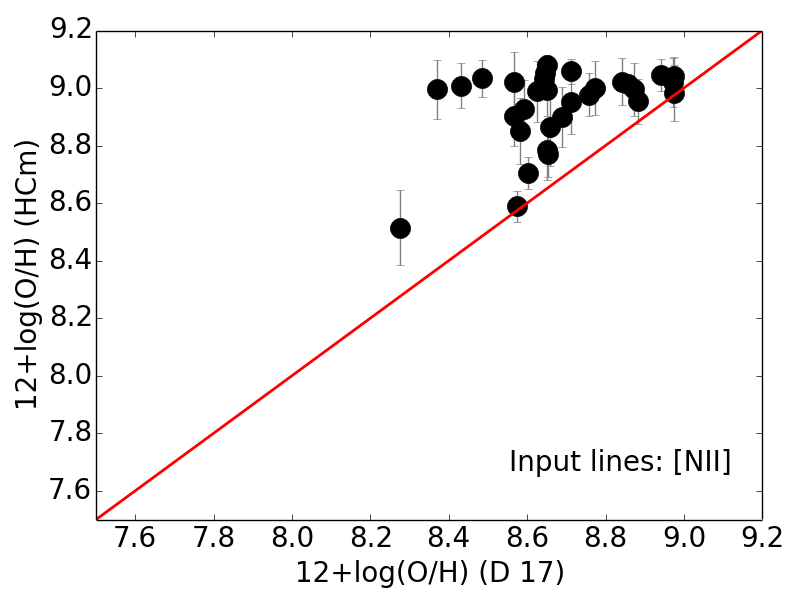} 
\includegraphics[width=0.49\textwidth]{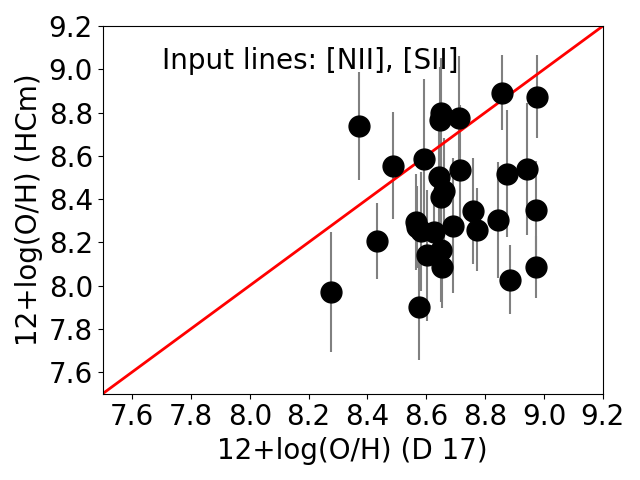}\\
\caption{Comparison between total oxygen abundances 12 + log(O/H) derived using the method described in this work (HCM) 
and those taken from \cite{dors17} from tailored photoionization models. 
\textit{Upper left}: comparison when all the lines are used.
\textit{Upper right}: comparison in the absence of [\oiii] $\lambda$ 4363 \AA\ and [\oii] $\lambda$ 3727 \AA. 
\textit{Middle left}: comparison when only  [\oii] $\lambda$ 3727 \AA\ and [\oiii] $\lambda$ 5007 \AA\ are included.
\textit{Middle right}: comparison when only [\oiii] $\lambda$ 5007 \AA\ and [\nii] $\lambda$ 6583 \AA\ are included.
\textit{Bottom left}: comparison when only [\nii] $\lambda$ 6583 \AA\ is provided.
\textit{Bottom right}: comparison when only [\nii] $\lambda$ 6583 \AA\ and [\sii] $\lambda$ 6717+6731 \AA\AA\ are included.
The solid line represents the 1:1 relation.
In the legend [\oii] stands for $\lambda$3727 \AA, 
\oiiin\ for $\lambda$5007 \AA, [\nii] for $\lambda$6583 \AA, and [\sii] for $\lambda$$\lambda$6717+6731 \AA\AA.}
\label{comp_oh}
\end{center}
\end{figure}

In Fig. \ref{comp_oh} we compare the total oxygen abundance derived for the control sample by \cite{dors17}
with those obtained by HCM  when all or only some 
of the input lines are used\footnote{More detailed comparisons including N/O and the ionization 
parameter can be found in \cite{epm19}.}.
The option of restricting the number of input lines simulates common observing conditions when only limited sensitivity 
or spectral coverage of the detector is available. The left upper panel shows the 
best case scenario when all possible emission lines are provided. There is a good agreement between both 
sets with a dispersion of 0.21 dex and a residual of -0.01 dex. The upper right panel displays the relation 
when lines [\oiii] $\lambda$ 4363 \AA\ and [\oii] $\lambda$ 3727 \AA\ are not included. This is common case 
when the blue part of the spectrum is not available (e.g. in the Sloan Digital Sky Survey at very low 
redshifts) and the [\oiii] $\lambda$ 4363 \AA\ is to faint to be observed. In this case, the dispersion is 
nearly the same but the residual increases by 0.1 dex. Even when only a couple of lines or only 
[NII] $\lambda$ 6583 \AA is available the agreement is good, with deviations from the abundances lower 
than the usual uncertainties. Table \ref{dispersion} shows the mean and standard deviation of the residuals 
of the comparison cases presented in Fig. \ref{comp_oh}.

\begin{table}
\label{dispersion}
\begin{center}
\caption{Mean and standard deviation of the residuals between the O/H from \cite{dors17} and 
the values derived by HCM using different input lines as shown in Figure \ref{comp_oh}. 
In the table [\oii] stands for $\lambda$3727 \AA, 
\oiiin\ for $\lambda$5007 \AA, [\nii] for $\lambda$6583 \AA, and [\sii] for $\lambda$$\lambda$6717+6731 \AA\AA.}
\begin{tabular}{clcc}
\hline
\hline
&  Input emission lines  &  Mean $\Delta$(O/H) & St.dev. $\Delta$(O/H) \\
\hline
&  All lines &  - 0.01  &  0.21  \\ 
& \oiiin, [\nii], [\sii]  &  +0.15  &  0.26 \\
& [\oii], \oiiin  &  - 0.11  &  0.21   \\
& \oiiin, [\nii]  &  -0.24  &  0.15 \\
& [\nii]  &  - 0.25   &   0.16     \\
& [\nii], [\sii]  &  + 0.29   &   0.29     \\
\hline
\end{tabular}
\end{center}
\end{table}

\subsection{Consistency with the direct method}

\begin{figure}
\begin{center}
\includegraphics[width=0.8\textwidth]{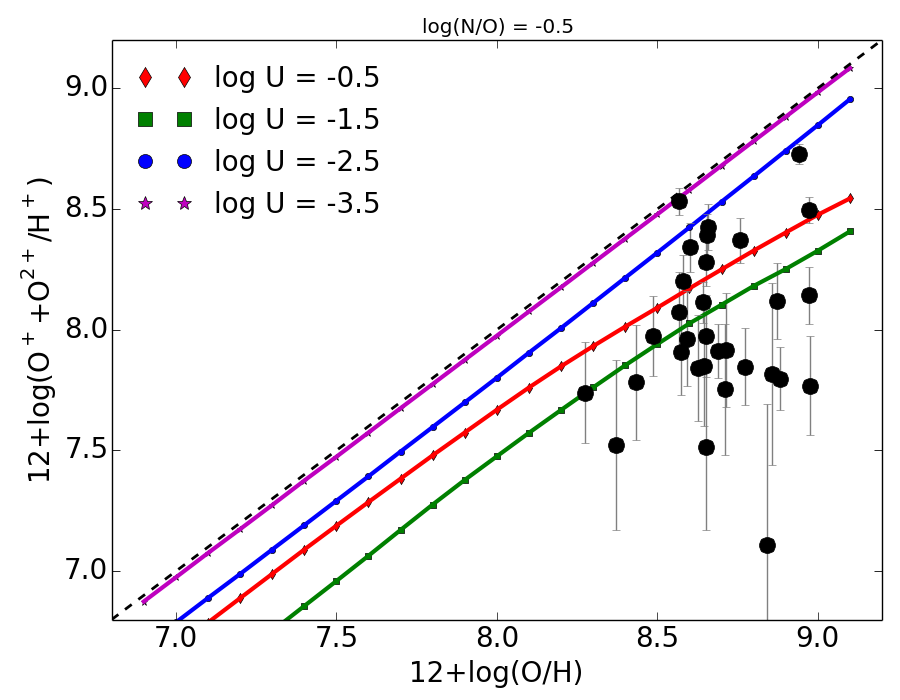} 
\caption{Comparison between total oxygen abundance and the addition of the abundances of the two
main oxygen ions O$^+$ and O$^{2+}$ observed in the optical part of the spectrum.
Models are represented using solid lines for different values of $U$.
Black circles represent the data from \cite{dors17} whose total abundances where calculated
using tailored models, while their ionic abundances were calculated following the \te\ method.
The dashed black line represents the 1:1 relation.}
\label{icf}
\end{center}
\end{figure}

There is known discrepancy between the chemical abundances derived using the \te\ method in NLRs 
of type-2 AGNs, leading to very low values if compared to those obtained from some photoionization 
models (e.g. \cite[Dors et al. 2015]{dors15}). The code HCM has proved to be in accordance with the 
\te\ method in star-forming regions (\cite[P\'erez-Montero 2014]). Thus, we can use HCM to 
investigate the possible origin of the discrepancies in AGNs.

In Fig. \ref{icf} we show the total oxygen abundance derived by \cite{dors17} for their sample of 
Sy2 galaxies, compared to the addition of the abundances of the most prominent oxygen ionic species in the 
optical part of the spectrum, i.e. O$^+$ and O$^{2+}$, calculated using the \te\ method.
The addition of the relative ionic abundances of the oxygen is 0.7 dex lower that the one derived 
by the models. Figure \ref{icf} shows also predictions from the grid of models for different 
ionization parameter values. The difference is well explained as an important dependence 
on the total metallicity and ionization parameter. This result highlights the importance of 
using models to derive the total oxygen abundance in NLRs of AGNs when only opital lines are 
available, as ionization correction factors (ICFs) are far to be negligible, contrary to 
star-forming regions.

\end{document}